\documentclass[12pt]{spieman}  
\usepackage{amsmath,amsfonts,amssymb}
\usepackage{graphicx}
\usepackage{setspace}
\usepackage{tocloft}

\title{Fabrication and characterization of optical micro/nanofibers}

\author[a]{Elaganuru Bashaiah}
\author[a]{Shashank Suman}
\author[a]{Resmi M}
\author[a]{Bratati Das}
\author[a,*]{Ramachandrarao Yalla}
\affil[a]{School of Physics, University of Hyderabad, Hyderabad, Telangana, India, 500046}

\cftpagenumbersoff{figure}
\cftpagenumbersoff{table} 
\begin{document} 
\maketitle

\begin{abstract}
We experimentally demonstrate the fabrication of optical micro/nanofibers (MNFs) using chemical etching and gas-flame techniques. In the chemical etching technique, a two-step process involves 40\% and 24\% of hydrofluoric acid solutions for the first and second steps, respectively. The measured diameters of MNFs range is 0.34 $\mu m$ - 1.4 $\mu m$. In the gas-flame technique, we design the pulling parameters in a four-step process to achieve the desired diameter of MNFs. The single-mode fiber is adiabatically tapered using high-precision stages while heating the fiber with hydrogen-oxygen flame. The measured diameters of pulled MNFs range is 0.48 $\mu \mathrm{m}$ - 0.53 $\mu \mathrm{m}$, showing good correspondence with the designed diameters. Due to the strong confinement of the field around the MNF, it has diverse applications in various fields, such as sensing, nanophotonics, quantum optics, quantum photonics, and nonlinear optics.
\end{abstract}


{\noindent \footnotesize\textbf{*}Ramachandrarao Yalla,  \linkable{rrysp@uohyd.ac.in} }

\begin{spacing}{1}  
\section{Introduction}
\label{sect:intro}
 Controlling quantum states of light and matter at the most fundamental level is the core focus of the future technology known as quantum technology \cite{1,2,3,4,lu2023quantum,heindel2023quantum,chang2023nanowire,luo2023recent,giordani2023integrated,kudalippalliyalil20243d}. This field encompasses a wide range of techniques that have been proposed and thoroughly explored through the use of micro/nanoscale dimensions. Numerous examples of these techniques include high-quality factor micro-structured resonators \cite{5}, diamond nano-beam cavities \cite{6}, silicon nitride alligator photonic crystal waveguides \cite{7}, photonic crystal cavities \cite{8}, plasmonic metal nano-structures \cite{9}, and sub-wavelength diameter tapered silica fibers \cite{10,11,12,13,14,elaganuru2024highly,resmi2023efficient,das2023efficient,resmi2024channeling,pathak2017fabrication,ascorbe2016high,resmi2024highly,kaur2021fabrication,wu2013optical,zhang2024optical,fang2024parallel,praveen2023particle,wang2024optical}. It is worth emphasizing that optical micro/nano-fibers (MNFs) have recently emerged as promising candidates among the above-mentioned micro/nano-structures due to their distinctive attributes, like the robust confinement of the electromagnetic field facilitated by the nanoscale proportions of the fibers. This strong confinement enables the efficient manipulation of quantum states of light and matter, allowing for the creation of quantum nodes. Furthermore, the efficient transmission lines between quantum nodes are facilitated by ultra-high light transmission through the guided modes of the MNF. Additionally, the automatic fiber integration feature of MNFs, which eliminates the need for additional optical elements, enables their seamless integration into a photonic network for quantum entanglement between two distinct nodes in a quantum network \cite{3}. It is worth noting that MNFs have garnered attention in various research fields, such as quantum optics \cite{lu2023quantum}, cavity quantum electrodynamics \cite{13}, nano-photonics \cite{heindel2023quantum,chang2023nanowire,luo2023recent}, and sensing \cite{10,11,12,13,14,15}. Chemically etched MNFs have been utilized for in situ characterization through scattering loss analysis \cite{suman2024situ}. Chemically etched optical nanofiber tips have been employed to couple fluorescence photons from quantum dots, making them valuable for quantum optics applications \cite{resmi2024channeling}. Additionally, chemically etched fiber tips have been widely used in humidity sensing \cite{ascorbe2016high}, concentration and refractive index measurements \cite{zaca2018etched}, and pH sensing \cite{pathak2017fabrication}. Pulled MNFs have been employed to efficiently channel fluorescence photons from single quantum dots into the guided modes of optical nanofibers \cite{11}. Additionally, cavity quantum electrodynamics on a nanofiber have been explored using a composite photonic crystal cavity \cite{13}. An optical interface is created by laser-cooled atoms trapped in the evanescent field surrounding an optical nanofiber \cite{10}, facilitating particle trapping with optical nanofibers \cite{praveen2023particle}.

Various techniques have been proposed and developed for the fabrication of MNFs. Examples include self-modulated taper drawing \cite{tong2003subwavelength, tong2005self}, flame-brushing and pulling \cite{brambilla2004ultra,brambilla2010optical,lee2019fabrication,hoffman2014ultrahigh},  CO$_2$ laser-assisted heating and pulling \cite{ward2006heat}, electrically heated taper drawing of glass fibers \cite{shi2006fabrication}, laser-heated pulling and bending \cite{lazarev2003formation}, mechanical polishing \cite{grosjean2007fiber}, and chemical etching \cite{kbashi2012fabrication, zaca2018etched,liao2010suspended,bashaiah2024fabrication,resmi2024channeling,suman2024situ}. Additionally, a combination of electric arc heated micro-pulling with chemical etching \cite{huo2006fabrication} and melt-stretching with chemical etching \cite{ren2007preparation} have been demonstrated experimentally. In the self-modulated taper drawing method \cite{tong2003subwavelength}, Tong \textit{et al}. demonstrated a tapered fiber through a two-step drawing process for crafting extended free-standing silica wires, achieving diameters as small as 50 nm. Flame-brushing and pulling is the primary methods employed for drawing MNFs from standard optical fibers. In this process, a hydrogen flame is utilized to heat the fiber. Applying a specific pulling force, the fiber undergoes gradual stretching and elongation, resulting in a reduced diameter until the desired length or the target diameter of the fiber taper is achieved. In this flame-brushing and pulling method, Brambilla \textit{et al} \cite{brambilla2004ultra} demonstrated the reproducibility of taper loss for a taper diameter of 750 nm. Hoffman \textit{et al} \cite{hoffman2014ultrahigh} fabricated ultrahigh transmission optical nanofibers with diameters of 520 nm.  In the CO$_2$ laser-assisted heating and pulling method, Ward \textit{et al} \cite{ward2006heat} demonstrated a fiber tapered to a diameter of 3-4  $\mu$m. Shi \textit{et al} \cite{shi2006fabrication} fabricated 650 nm tapered fiber in the electrically heated taper drawing method. In the laser-heated pulling and bending technique, the creation of a fiber tip with a short tapering length of approximately 300 $\mu$m and a diameter of about 50 nm was demonstrated. In the chemical etching technique, Kbashi \textit{et al} \cite{kbashi2012fabrication} fabricated tapered fiber with a diameter of 357 $nm$. Zaca \textit{et al} \cite{zaca2018etched} also employed a chemical etching technique to fabricate a tapered fiber with a diameter of 7.3 $\mu$m.

In this paper, we experimentally demonstrate the fabrication of MNFs using chemical etching and gas-flame techniques. Hydrofluoric (HF) acid is employed as the etching agent in the chemical etching technique. A two-step etching technique involves 40\% and 24\% of HF acid solutions for the first and second steps, respectively. The measured diameters of MNFs are from 0.34-1.4 $\mu m$. In the gas-flame technique, we design the pulling parameters in a four-step process to achieve the desired diameter of MNFs. The single-mode fiber (SMF) with an outer diameter of 125 $\mu$m is adiabatically tapered using high-precision stages while heating the fiber with hydrogen-oxygen flame. The measured diameters of MNFs range is 0.48-0.53 $\mu \mathrm{m}$, showing good correspondence with the designed diameters.

\section{Fabrication and characterization procedures}
\label{sec:Exp}
\subsection{Pre-preparation}

We perform experiments in a clean booth as dust or contamination substantially threatens the successful execution of these MNFs. A 40 cm long standard SMF (1550B-HP, Coherent) is utilized in the fabrication process. This SMF consists of a 9 $\mu m$ core, a 125 $\mu m$ clad, and a 250 $\mu m$ outer jacket. The outer jacket at the center of the SMF is stripped using a three-hole fiber stripping tool (FTS4, Thorlabs). Alternatively, the center of the SMF is immersed in acetone (10020LC250, Finar Ltd.) for a duration ranging from 15-20 minutes. This process reduces fiber diameter from 250 $\mu m$ to 125 $\mu m$ over a $\approx$ 3 cm length. The fiber is cleaned with clean room wipes. Such a cleaned SMF is used for further experiments. The fabrication is performed in a fume hood, which vents the fumes from the acid away from the surroundings. We use gloves, goggles, respiratory masks, aprons, and shoes that are designed to deal with hazardous chemicals during the experiment. Residual HF is disposed of in a hazardous chemicals management facility.

\subsection{Etching process}

\begin{figure}[!h]
	\centering
	\includegraphics[width=\linewidth]{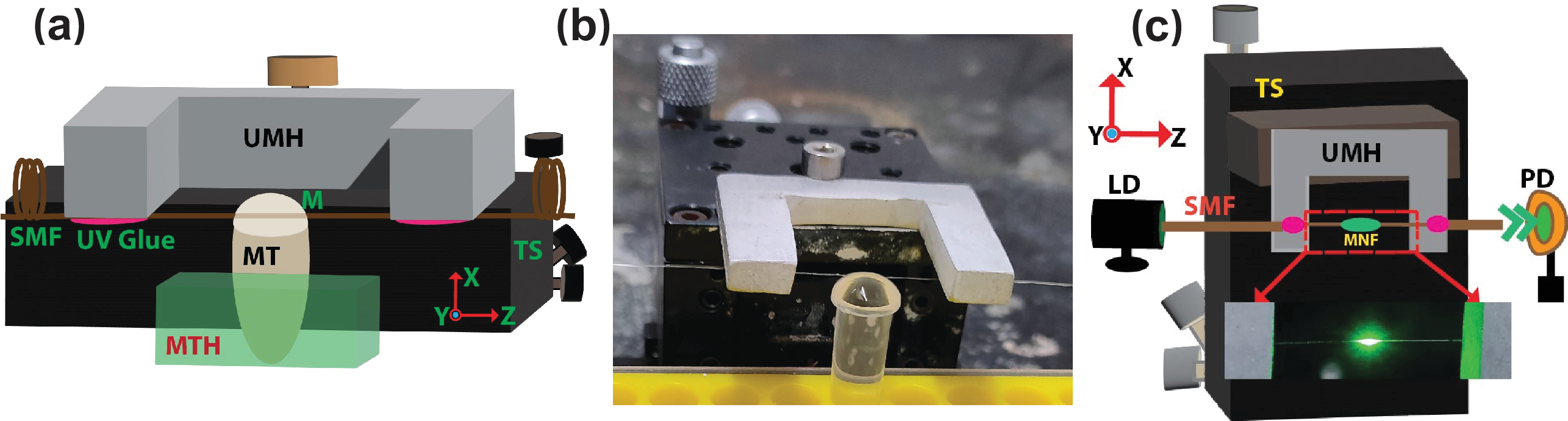}
	\caption{(a) The schematic for the fabrication of micro/nanofibers (MNFs) using a chemical etching technique. (b) The photo of the experimental setup for the fabrication of the MNF. (c) The schematic of the experiment for optical transmission measurement of MNFs. The inset shows the scattering profile.}
	\label{fig1}
\end{figure}

The SMF is fixed to the in-house custom-designed U-shaped metal holder (UMH) on both sides of surfaces using ultraviolet (UV) curable glue, cured by UV light for approximately 20 minutes. It is crucial to ensure that the stripped section of the SMF is positioned between the two sides of the UMH. A schematic of the experimental setup for the fabrication is shown in Fig. \ref{fig1} (a). Several essential tools include the SMF, UMH, UV curable glue, translation stage, TS (TS 65, Holmarc), microcentrifuge tube (MT), meniscus (M), and microcentrifuge tube holder (MTH). The UMH is placed on the TS, and it is secured with a cap screw to translate in the \textit{xyz} directions. A double-sided tape is attached between the table and MTH for stability. The double-sided tape between the table and the TS stabilizes the experimental setup and prevents external disruptions. The UMH is placed just above the 500 $\mu l$ MT so the fiber can be immersed in the meniscus. Once the setup is arranged, HF acid is carefully poured into the MT until the meniscus forms at the top.

This technique utilizes the meniscus on top of the 500 $\mu l$ MT due to the HF solution's surface tension. Subsequently, the SMF is placed within the HF acid's upper meniscus using the TS. A two-step etching technique involves controlling the etching speed of HF acid. The original concentration of 40\% HF acid (Avra, ASH2565) is poured into the 500 $\mu l$ MT with the help of a micropipette. The SMF is placed in the upper meniscus of HF acid, and its position is precisely controlled by TS in three directions. Here, the SMF contacting area of the meniscus is around 1.3 cm in length. The etching time is 65-70 minutes. In the second step, 40\% HF acid is diluted to 24\% using de-ionized water to control the etching speed. The 24\% HF acid is poured in the 500 $\mu l$ MT with the micropipette. A tapered SMF is positioned so that it just makes contact with the upper meniscus for a length of 3-4 mm. The tapered SMF diameter is etched by 24\% HF acid in 25–30 minutes. The corresponding photo of the chemical etching experiment is shown in Fig. \ref{fig1} (b). One can readily see that fiber is placed in the upper meniscus. The other mentioned experiment components can also be seen.
\subsubsection{Optical characterization}

The experimental sketch for the optical characterization of the MNF is shown in Fig. \ref{fig1} (c). We measure the power transmitted through its guided mode. The fabricated MNFs are safely attached to another TS with the help of UMH. The double-sided tape is placed between the UMH and TS. A laser diode, LD (DL-G-5, Holmarc) at 532 nm is used for the measurements. A single-mode fiber patch cable along with a fiber connector cum terminator (FCT) (B30126A9 and BFT1, Thorlabs) is used to guide the laser into the MNF. Before we connect the FCT to the MNF, both the ends of the MNF are stripped using a three-hole fiber stripping tool, and it is cleaved perpendicularly using a fiber cleaver (XL411, Thorlabs). One end of the MNF is connected to the FCT, and another is placed in front of a photodiode. The optical power measurements are carried out using the photodiode (PD) (S120VC, Thorlabs) connected to a power meter console (PM100A, Thorlabs). The output power is measured for 15 minutes. The fiber is cleaved perpendicularly before the MNF region for measuring the input power acurately. The optical transmission characteristic is derived, and the transmission ($T$) is calculated as $\frac{o/p}{i/p} \times 100$. The inset shows a typical photo of the MNF while sending the laser through it. One can readily observe that a significant field is outside the MNF. Such optical characterized MNFs are safely transferred to a metal holder for morphological characteristics. We placed the metal holder below the UMH. The MNF position is controlled using the TS. Then, MNF samples were fixed on the surface of the metal holder using UV-curable glue for 10 minutes.

\subsection{Pulling process} 

\begin{figure}[!h]
	\centering
	\includegraphics[width=\linewidth]{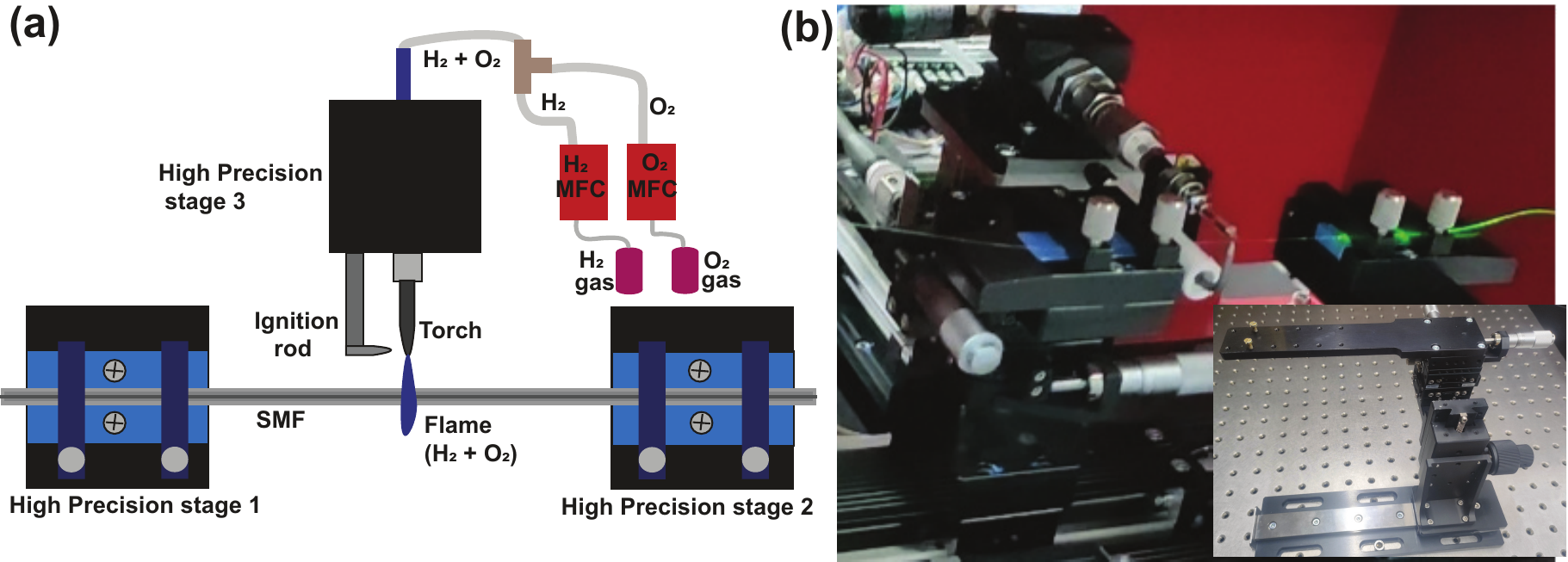}
	\caption{(a) The schematic for the fabrication of an optical micro/nanofiber (MNF) using gas flame technique. (b) The picture of the experimental setup for the fabrication of the MNF.}
	\label{fig2}
\end{figure}

A model has been developed to design and predict the diameter profile of the MNF by ensuring the maintenance of the adiabatic tapering condition along the entire length of the fiber axis \cite{birks1992shape}. Three computer-controlled high-precision stages are used in the gas-flame technique (Deltafiber, TFEX-2810). Two of these stages are used to pull the fiber, while the third controls the position of the microheater. The stages have a 10 $\mu \mathrm{m}$ resolution for $x$ and $y$ movements and a $\pm$12 mm travel range in each direction. For $z$ movement, the resolution is 0.02 $\mu \mathrm{m}$, with a maximum movement of 160 mm, including fiber swing. The design and development of a micro-heater is a pivotal step in achieving a stable and consistent flame. A micro nozzle to generate a steady and laminar flow of a gas mixture comprising hydrogen and oxygen is employed as depicted in Fig. \ref{fig2}. The nozzle size is approximately $0.5 \ \mu \mathrm{m}$ with an array of three nozzles horizontally. The heater position is controlled with precision using computer-controlled a high-precision stage. The establishment of a micro-flow technique is also a crucial step in achieving a uniform gas flame. The high-precision mass flow controllers for regulating the flow of oxygen and hydrogen are employed, as shown in Fig. \ref{fig2}. The volume ratio of hydrogen to oxygen and flow rates is optimized to achieve a stable and laminar flow. The volume ratio and flow rate are maintained at 1:0.28 and $115\  \mathrm{SCCM}\left(\mathrm{H}_2 \ 90 \ \mathrm{SCCM}\right.$ and $\left.\mathrm{O}_2 \ 25 \ \mathrm{SCCM}\right)$, respectively.
 
The primary objective of the pre-alignment procedure is to ensure uniform tension along the fiber. We position a cleaned SMF in the $v$-grooves of magnetic fiber clamps on high-precision stages, as conceptually illustrated in Fig. \ref{fig2} (a). We tension the fiber by gradually moving the pulling stages apart until the fiber slides through the fiber clamps. While moving, we imaged the fiber through the camera placed perpendicular to the fiber to ensure the fiber reached a uniform tension \cite{hoffman2014ultrahigh}. The micro-heater is positioned such that the fiber is approximately $1 \mathrm{~mm}$ above it using high-precision stages. The fiber's position along the flow direction is optimized to place it at the center of the hot zone produced by the micro-heater. Once the fiber and nozzle are aligned, the fiber is ready for pulling. High-precision stages are systematically employed to pull the fiber. The parameters for pulling are controlled using software. Four main parameters - swing width, swing speed, stretch speed, and stretch length - determine the scan length $(L)$ and stretching distance $(z)$.

\subsubsection{Optical characterization}
During the fabrication process, the system is constantly monitored for optical transmission through the guided modes of the MNF, as shown in Fig. \ref{fig2} (b). This measurement procedure is the same as discussed in the preceding section. The recorded transmission signal is subsequently analyzed to determine the optical signal transmission value while pulling the MNF. It is important to note that the system can measure the optical transmission with a high resolution. Such optically characterized MNFs are safely transferred to a metal holder for morphological characteristics, as shown in the inset of Fig \ref{fig2} (b). We placed the metal holder below the high-precision stages. The MNF position is controlled using a three-axis stage. Then, MNF samples were fixed on the surface of the metal holder using UV-curable glue.

\subsection{Morphological characterization}

A field emission scanning electron microscope (FESEM) is used to measure the diameter profile of the MNF accurately. To avoid charge-up on the MNF, a mixture of gold and palladium is sputter-coated on the samples for 3 minutes. Subsequently, it is carefully placed on the metal holder housing the MNF within the vacuum chamber of the FESEM. We acquire MNF images along the fiber axis with an evenly distributed spacing of 100 $\mu$m. We measure the diameters of the MNF from one end to the other. By plotting the obtained data, we construct the MNF diameter profile as a function of the distance along the fiber axis. We define the center at the minimum and most uniform diameter along the fiber axis. We use statistical analysis techniques to assess the diameter distribution, deriving parameters such as the mean diameter and standard deviation to gain insights into the uniformity of the MNF diameter.

\section{Results and Discussion}

\begin{figure}[!h]
	\centering
	\includegraphics[width= 8cm]{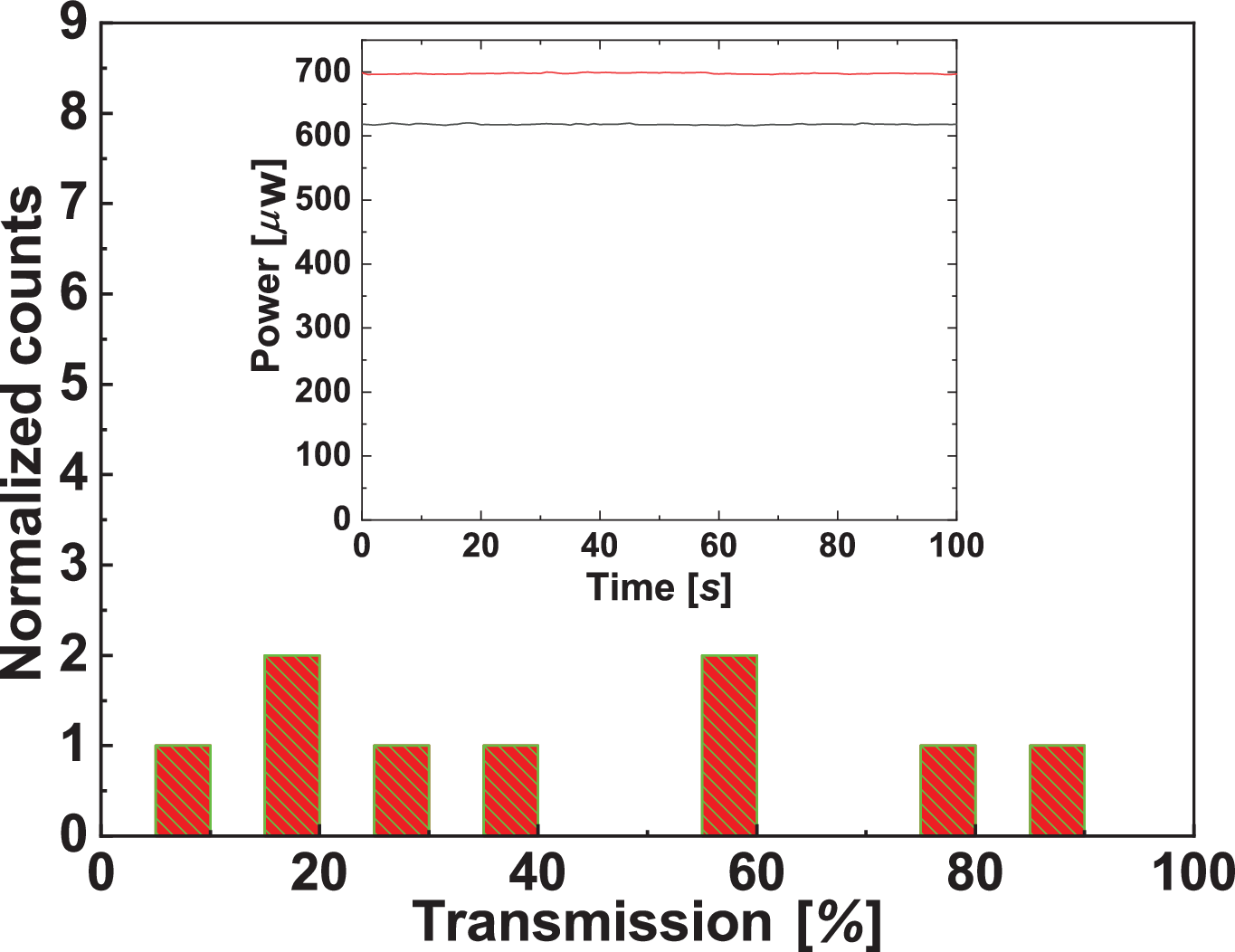}
	\caption{The measured histogram profiles for the optical transmission of chemically etched MNFs.}
	\label{fig3}
\end{figure}

We performed transmission measurements for nine chemically etched fibers. A summary of such measurements is plotted as a histogram, as shown in Fig. \ref{fig3}. The horizontal axis corresponds to the transmission percentage, and the vertical axis corresponds to the normalized counts. The inset shows typical input (red) and output (black) power traces as a function of time. The measured optical transmission ranges from a minimum of 5.0 ($\pm$0.5)\% to a maximum of 88 ($\pm$9)\%. The maximum occurrence of transmission is at 52 ($\pm$4)\% and 18 ($\pm$2)\%, corresponding measured MNF diameters are 6.73 $\mu m$ and 1.87 $\mu m$, respectively. As seen in the inset of Fig. \ref{fig1} (c), the observed scattering loss is significant for etched fibers compared to pulled fibers. This may be attributed to the non-adiabatic tapering of the MNF and the lack of smoothness on its surface. One can readily see that most of the light is scattered outside the fiber. The measured transmission varies widely as seen in Fig. \ref{fig3}. The maximum occurrence of tramsmision is at 52 ($\pm$4)\% and 18 ($\pm$2)\%, corresponding measured MNF diameters are 6.73 $\mu m$ and 1.87 $\mu m$, respectively. One can readily see that thinner MNFs have low transmission compared to thicker fibers. This is due to the surface roughness being predominant for the thinner fibers. This is because thinner fibers have a higher surface roughness due to chemical etching.
 
\begin{figure}[!h]
	\centering
	\includegraphics[width=0.8\linewidth]{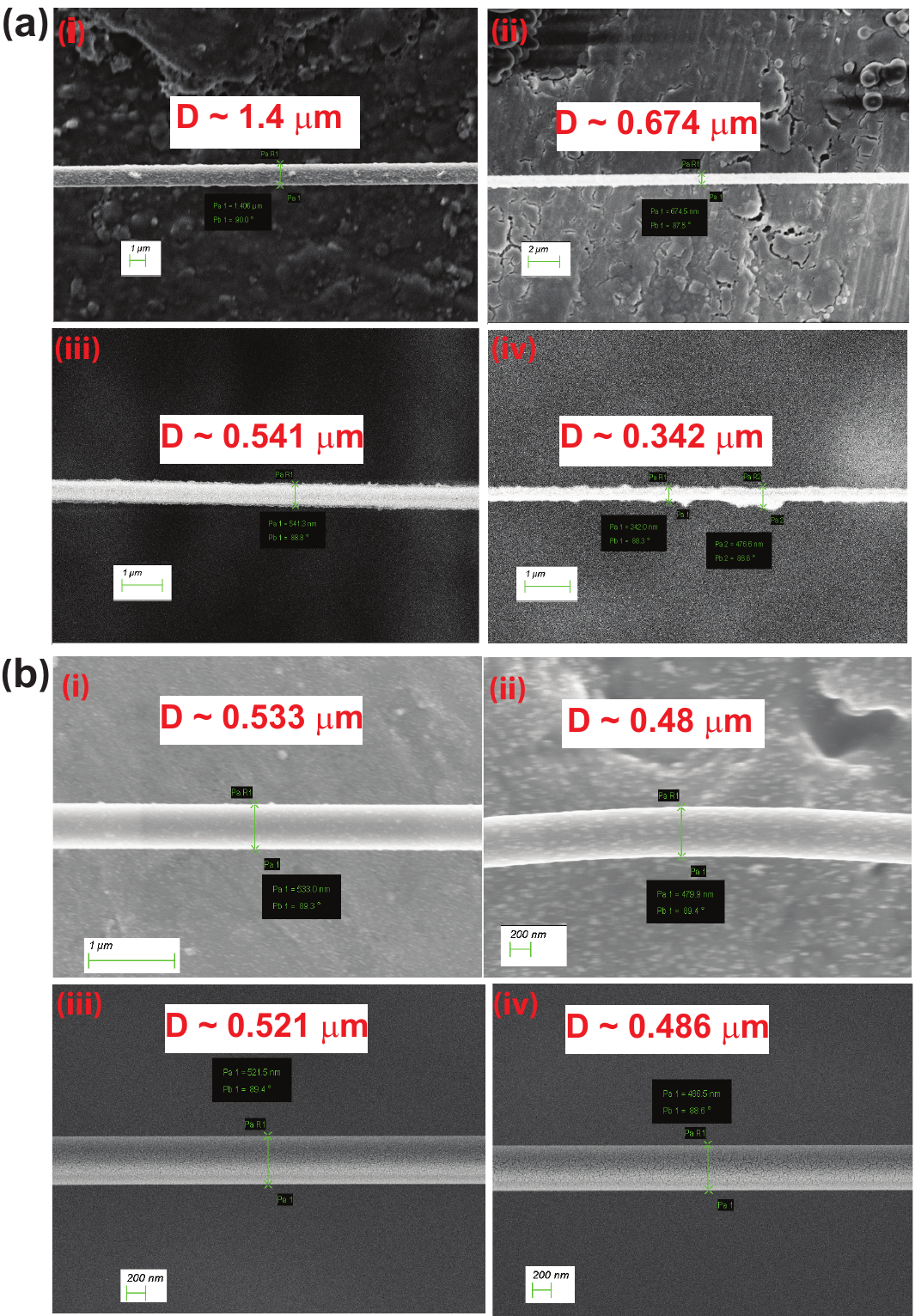}
	\caption{(a) Typical images of chemically etched micro/nanofibers showing in diameters (D) (i) 1.40 ($\pm$0.08) $\mu m$, (ii) 0.674 ($\pm$0.04) $\mu m$, (iii) 0.541 ($\pm$0.03) $\mu m$, and (iv) 0.342 ($\pm$0.04) $\mu m$. (b) Typical images of mechanically pulled micro/nanofibers showing in diameter (D) (i) 0.533 ($\pm$0.02) $\mu m$, (ii) 0.480 ($\pm$0.02) $\mu m$, (iii) 0.521 ($\pm$0.02) $\mu m$, and (iv) 0.486 ($\pm$0.02) $\mu m$.}
	\label{fig4}
\end{figure}

We fabricated nine MNF samples using the chemical etching technique and measured four of them. Typical FESEM images of the four etched MNFs are shown in Figs. \ref{fig4} (a). The measured MNFs diameter of (i) 1.4 ($\pm$0.08) $\mu m$, (ii) 0.674 ($\pm$0.04)  $\mu m$, (iii) 0.541 ($\pm$0.03)  $\mu m$, and (iv) 0.342 ($\pm$0.04) $\mu m$. Similarly, we fabricated two MNF samples using the gas flame technique and measured two MNF samples of the mechanically pulled fibers, as shown in Figs. \ref{fig4} (b). The diameters shown in Fig. \ref{fig4} (b) are (i) 0.533 ($\pm$0.02) $\mu m$ and (ii) 0.480 ($\pm$0.02) $\mu m$, which belong to one sample at different locations, while the diameters in Fig. \ref{fig4} (b) (iii) 0.521 ($\pm$0.02) $\mu m$ and (iv) 0.486 ($\pm$0.02) $\mu m$ belong to another sample at different locations. We also measured the diameters of the MNFs at different tapered regions for both samples.

As seen in Figs. \ref{fig4} (a) (i)-(iv), the measured MNF diameters vary from 0.342 $\mu m$ to 1.420 $\mu m$ for chemically etched fibers. As seen in Figs. \ref{fig4} (b) (i)-(iv), the measured MNF diameters vary from 0.486 $\mu m$ to 0.581 $\mu m$ for pulled fibers. MNFs with diameters of a few hundred nanometers to micrometers have been realized in both methods. One can readily see that MNF surfaces are very smooth for pulled fibers compared to etched fibers. This may be due to the etching process, which results in a wide range of diameters. To fabricate chemically etched MNFs, we optimized the etching rate by controlling the concentration of HF. Also, the etching rate depends on the temperature and humidity. The measured temperatures are from 32$^o$C to 39$^o$C, and the humidity is from 18-44\%. The optimized etching times are 65-70 minutes for the first step and  25-30 minutes for the second step, respectively. In the first step, the fiber diameter is reduced from 125 $\mu m$ to 15-20 $\mu m$.  In the second step, the fiber diameter reduces from 15 $\mu m$ to 0.5 $\mu m$, the corresponding etching rate of 0.3-0.4 $\mu m$ per minute. Two-step etching is preferred over single-step etching for its ability to precisely control the etching rate, which is crucial for achieving MNFs. In the first step, using 40\% HF nearly removes the cladding at an etching rate of 1.5-1.6 $\mu m$ per minute. If this concentration were used for the entire process, it would be challenging to control the MNF size, and the resulting taper would be too sharp, impacting transmission loss. Conversely, using 24\% HF throughout the entire process would significantly lengthen the etching time, leading to changes in environmental conditions and potential inconsistencies in diameter. Previous studies also support the advantages of two-step etching in this context \cite{kbashi2012fabrication}. A higher concentration of HF can be used in the first step as it does not require precise control to etch the cladding, but the concentration in the second step is crucial.

\begin{figure}[!h]
	\centering
	\includegraphics[width=\linewidth]{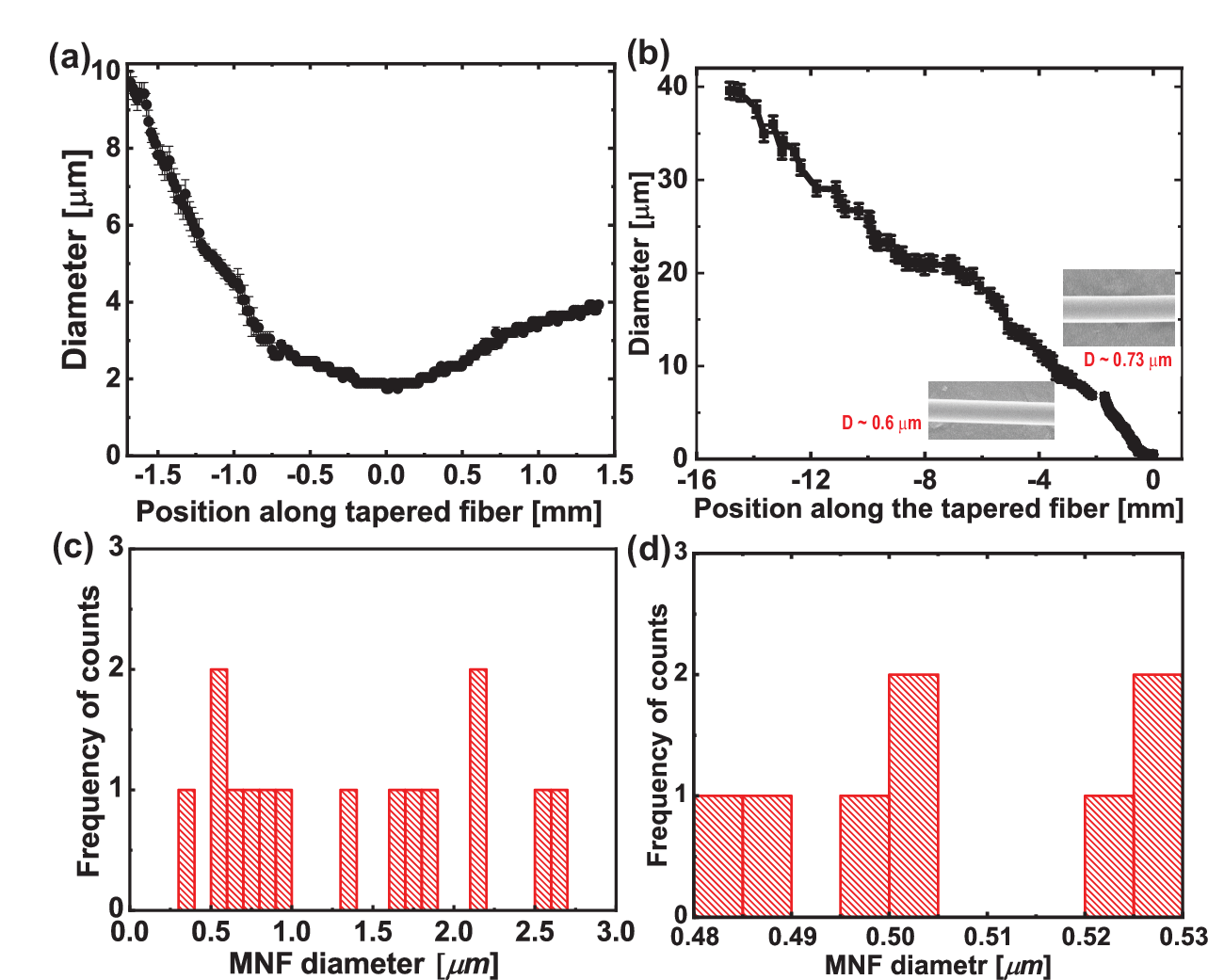}
	\caption{(a) and (b) are typical diameter profiles of the MNF for etched and pulled fiber, respectively. (c) and (d) are histogram profiles for the sizes of the MNFs for etched and pulled fibers, respectively.}
	\label{fig5}
\end{figure}

To fabricate mechanically pulled MNFs, we designed and optimized the pulling parameters, such as stretch distance and scan length \cite{birks1992shape}. The optimum parameters are: In the first step, the fiber oscillates with a swing width of $7 \ \mathrm{mm}$ and a swing speed of $3.22 \ \mathrm{mm} / \mathrm{s}$ in the flame, with no stretching ($z_{1^{-}}$ value is 0). This step is solely for melting the fiber to its melting point. In the second step, the fiber is stretched with a stretch speed of $0.138\  \mathrm{mm} / \mathrm{s}$, while oscillating with a swing width of 7 $\mathrm{mm}$ and a swing speed of $3.22\  \mathrm{mm} / \mathrm{s}$. The stretching distance per cycle is about $0.15\  \mathrm{mm}$ and remains constant across the steps. This results in 40 cycles to achieve a stretching distance of $\left(z_2\right)$ of $6\  \mathrm{mm}$. In the third step, the fiber is stretched with a stretch speed of $0.12\  \mathrm{mm} / \mathrm{s}$, while oscillating with a swing width of $8\  \mathrm{mm}$ and a swing speed of $3.2\  \mathrm{mm} / \mathrm{m}$s. This involves 156 cycles to achieve a stretching distance of $\left(z_3\right)$ of $29.4\  \mathrm{mm}$. In the fourth step, the fiber is stretched with a stretch speed of $0.48\  \mathrm{mm} / \mathrm{s}$, while oscillating with a swing width of 2 $\mathrm{mm}$ and a swing speed of $3.2\  \mathrm{~mm} / \mathrm{s}$. This involves 98 cycles to achieve a stretching distance of $\left(z_{\mathrm{4}}\right)$ of $44\  \mathrm{~mm}$. Note that this length includes both sides of the MNF. Therefore, one side of the tapered fiber length is $22\  \mathrm{~mm}$. This process results in the creation of the MNF, located in the mid-section of the tapered optical fiber, where the core essentially disappears, and the original fiber's cladding becomes the core of the MNF. One can readily see that the measured diameter ($\sim$0.5 $\mu m$) is closely matched with the predicted diameter.

Figures \ref{fig5} (a) and (b) show typical measured diameter profiles of the MNF for the etched and pulled fiber, respectively. The horizontal axis corresponds to the position along the fiber axis, and the vertical axis corresponds to the MNF diameter. The measured etched MNF diameter ranges from 10 $\mu m$ to 2 $\mu m$, and the tapered length of the fiber is about 1.75 mm. The measured pulled MNF diameter ranges from 40 $\mu m$ to 0.5 $\mu m$, and the tapered length is about 15 mm. Typical tapered fiber diameters are shown in the inset of Fig. \ref{fig5} (b). As seen in Fig. \ref{fig5} (a), the tapered length (1.75 mm) of the etched fiber is shorter than the tapered length (15 mm) for pulled fiber, as shown in Fig. \ref{fig5} (b). The diameter profile of the tapered fiber was measured on both sides from the center for chemically etched MNFs. The diameter profile was measured only on one side from the center for the pulled fibers, as it is expected to exhibit symmetric behavior on the other side. Note that a typical overlap area of the SMF with HF droplet is in order of a few mm, which results in a short tapering length. We designed the tapering length for pulled fibers to be about 20 mm. The experimentally measured tapered length is about 15 mm. This discrepancy is because the measured maximum fiber diameter is up to 40 $\mu m$ only. Due to a technical difficulty in our setup, we could not measure up to 125 $\mu m$ \cite{hoffman2014ultrahigh}. Therefore, the tapering length would be more than 15 mm. We designed a long tapering length to maintain the adiabatic condition to see a high transmission through the MNFs.       

Similarly, we performed diameter measurements for fifteen etched samples. A summary of a minimum MNF diameter is plotted as a histogram, as shown in Fig. \ref{fig5} (c). The horizontal axis corresponds to the MNF diameter, and the vertical axis corresponds to the frequency of counts. The measured diameter ranges from 0.342 $\mu m$ to a maximum of 2.66 $\mu m$. The diameters of 0.5-0.6 $\mu m$ and 2.1-2.2 $\mu m$ were observed to be more probable, distinguishing them from the other samples. We also measured the diameters of two samples of pulled fibers. A summary of a minimum diameter is displayed in the histogram shown in  Fig. \ref{fig5} (d). The horizontal axis corresponds to the MNF diameter, and the vertical axis corresponds to the frequency of counts. The measured diameter ranges from a minimum of 0.48 $\mu m$ to a maximum of 0.53 $\mu m$. The diameter of 0.5-0.53 $\mu m$ is observed to be more probable, distinguishing them from the other samples. Note that the bin widths of histograms are chosen according to variations in the MNF diameters. As seen in Fig. \ref{fig5} (c), the histogram measurements for the diameters of MNFs range from 0.342 $\mu m$ to 1.4 $\mu m$ for the etched fiber. In contrast, as seen in \ref{fig5} (d), the histogram measurements for the diameters of MNFs range from 0.48 $\mu m$ to 0.53 $\mu m$ for the pulled fiber. One can readily see that the uniformity in diameters for the pulled fibers is better than etched fibers. The reproducibility in MNF diameter is far better for the pulling technique than the etching technique \cite{hoffman2014ultrahigh}. The measured minimum diameters were 0.486 ($\pm$0.02) $\mu m$ for MNF sample 1 and 0.480 ($\pm$0.02) $\mu m$ for MNF sample 2. We observed an error of 20 nm for both samples.

Previously, Kbashi et al. \cite{kbashi2012fabrication}. and Zaca et al. \cite{zaca2018etched}. bent the fiber and dipped it in HF acid to create tapered fibers, but this method often led to the fiber breaking during fabrication. To avoid this, we developed a method that forming a meniscus on the top of a microcentrifuge tube helps prevent the fiber from breaking. The etching is a simple and low-cost technique for fabricating MNFs compared to other techniques. However, we observed 0.342 $\mu m$ minimum diameter of the MNF with optical transmission of below 1\%. For 2 $\mu m$ diameter of MNFs, we measured 30-40\% optical transmission. The advantage of chemically etched MNFs is short tapered fibers compared to heat and pulled fibers. The long tapered fibers have handling issues while experimenting. The disadvantage of chemically etched MNFs is that optical transmission goes low when the tapered fiber diameter reaches less than 1 $\mu$m. To overcome these transmission issues, the pulling technique is established for low-loss MNFs \cite{tong2003subwavelength,brambilla2004ultra}, and it can maintain adiabatic tapering. Hence, the MNF will find more applications in quantum optics, quantum photonics \cite{nayak2018nanofiber,politi2009integrated}, and biomedical \cite{hale1994demonstration,pace2004refractive}. Regarding nanophotonics application, chemically etched MNF has been used by us for in situ characterization of them using scattering loss analysis \cite{suman2024situ}. Chemically etched optical nanofiber tip has been used to couple fluorescence photons from quantum dots, making them useful in quantum optics applications \cite{resmi2024channeling}. Chemically etched fiber tips have been widely used for humidity sensing \cite{ascorbe2016high}, concentration and refractive index \cite{zaca2018etched}, and pH sensing \cite{pathak2017fabrication}.

\section{Conclusion}
\label{sec:Con}
We experimentally demonstrated the fabrication of optical micro/nanofibers (MNFs) using chemical etching and gas-flame techniques. Hydrofluoric (HF) acid was employed as the etching agent, and an upper meniscus formed on the top of the microcentrifuge tube. A two-step etching technique involved 40\% and 24\% of HF acid solutions for the first and second steps, respectively. We observed the diameters of MNFs from 0.34-1.4 $\mu m$. The SMF is adiabatically tapered in the gas-flame technique using the high-precision stages. A four-step process has been devised to achieve an MNF diameter of approximately 0.5 $\mu \mathrm{m}$. The measured MNF diameter was in good agreement with the designed diameters within experimental errors. Due to the strong confinement of the field around the MNF, it has diverse applications in various fields, such as sensing, nanophotonics, quantum optics, quantum photonics, and nonlinear optics.

% \disclosures 
\subsection*{\textbf{Disclosures}}
The authors declare no conflict of interests.

\subsection* {\textbf{Code, Data, and Materials Availability}} 
Data may be obtained from authors upon reasonable request.

\subsection* {\textbf{Acknowledgments}}
Some part of the work was carried out in the wet chemistry lab under UGC-NRC in the School of Physics, University of Hyderabad. SS acknowledges funding support for the Chanakya-PG fellowship from the National Mission on Interdisciplinary Cyber-Physical Systems, of the Department of Science and Technology, Govt. of India through the I-HUB Quantum Technology Foundation (File No. I-HUB/PGF/2022-23/01). RM acknowledges the University Grants Commission (UGC) for the financial support (Ref. No.:1412/CSIR-UGC NET June 2019). BD acknowledges support from the Council of Scientific and Industrial Research (CSIR), India, for the fellowship (Ref. No.:09/0414(11066)/2021-EMR-I). RRY acknowledges support from the Scheme for Transformational and Advanced Research in Science (STARS) grant from the Ministry of Human Resource Development (MHRD) (File No. STARS/APR2019/PS/271/FS), Institute of Eminence (IoE) grant at the University of Hyderabad (File No. UoH-IoE-RC2-21-019), and the Council of Scientific and Industrial Research (CSIR) from the Human Resources Development Group (HRDG) (File No. 03/1487/2023/EMR-II).
%%%%% References %%%%%

\bibliography{references}   % bibliography data in report.bib

\begin{thebibliography}{10}

\bibitem{1}
M.~A. Nielsen and I.~L. Chuang, ``Quantum information and quantum
  computation,''  (2000).

\bibitem{2}
T.~D. Ladd, F.~Jelezko, R.~Laflamme, {\em et~al.}, ``Quantum computers,'' {\em
  nature} {\bf 464}(7285), 45--53  (2010).

\bibitem{3}
H.~J. Kimble, ``The quantum internet,'' {\em Nature} {\bf 453}(7198),
  1023--1030  (2008).

\bibitem{4}
H.~J. Kimble, Y.~Levin, A.~B. Matsko, {\em et~al.}, ``Conversion of
  conventional gravitational-wave interferometers into quantum nondemolition
  interferometers by modifying their input and/or output optics,'' {\em
  Physical Review D} {\bf 65}(2), 022002  (2001).

\bibitem{lu2023quantum}
Y.~Lu, A.~Sigov, L.~Ratkin, {\em et~al.}, ``Quantum computing and industrial
  information integration: A review,'' {\em Journal of Industrial Information
  Integration} , 100511  (2023).

\bibitem{heindel2023quantum}
T.~Heindel, J.-H. Kim, N.~Gregersen, {\em et~al.}, ``Quantum dots for photonic
  quantum information technology,'' {\em Advances in Optics and Photonics} {\bf
  15}(3), 613--738  (2023).

\bibitem{chang2023nanowire}
J.~Chang, J.~Gao, I.~Esmaeil~Zadeh, {\em et~al.}, ``Nanowire-based integrated
  photonics for quantum information and quantum sensing,'' {\em Nanophotonics}
  {\bf 12}(3), 339--358  (2023).

\bibitem{luo2023recent}
W.~Luo, L.~Cao, Y.~Shi, {\em et~al.}, ``Recent progress in quantum photonic
  chips for quantum communication and internet,'' {\em Light: Science \&
  Applications} {\bf 12}(1), 175  (2023).

\bibitem{giordani2023integrated}
T.~Giordani, F.~Hoch, G.~Carvacho, {\em et~al.}, ``Integrated photonics in
  quantum technologies,'' {\em La Rivista del Nuovo Cimento} {\bf 46}(2),
  71--103  (2023).

\bibitem{kudalippalliyalil20243d}
R.~Kudalippalliyalil, T.~Chakraborty, T.~E. Murphy, {\em et~al.}, ``3d
  self-aligning, polarization-independent fiber-to-chip couplers,'' in {\em
  Optical Fiber Communication Conference},  M1J--1, Optica Publishing Group
  (2024).

\bibitem{5}
K.~J. Vahala, ``Optical microcavities,'' {\em nature} {\bf 424}(6950), 839--846
   (2003).

\bibitem{6}
B.~J.~M. Hausmann, B.~J. Shields, Q.~Quan, {\em et~al.}, ``Coupling of nv
  centers to photonic crystal nanobeams in diamond,'' {\em Nano letters} {\bf
  13}(12), 5791--5796  (2013).

\bibitem{7}
S.-P. Yu, J.~Hood, J.~Muniz, {\em et~al.}, ``Nanowire photonic crystal
  waveguides for single-atom trapping and strong light-matter interactions,''
  {\em Applied Physics Letters} {\bf 104}(11)  (2014).

\bibitem{8}
E.~Yablonovitch, ``Inhibited spontaneous emission in solid-state physics and
  electronics,'' {\em Physical review letters} {\bf 58}(20), 2059  (1987).

\bibitem{9}
A.~Akimov, A.~Mukherjee, C.~Yu, {\em et~al.}, ``Generation of single optical
  plasmons in metallic nanowires coupled to quantum dots,'' {\em Nature} {\bf
  450}(7168), 402--406  (2007).

\bibitem{10}
E.~Vetsch, D.~Reitz, G.~Sagu{\'e}, {\em et~al.}, ``Optical interface created by
  laser-cooled atoms trapped in the evanescent field surrounding an optical
  nanofiber,'' {\em Physical review letters} {\bf 104}(20), 203603  (2010).

\bibitem{11}
R.~Yalla, F.~Le~Kien, M.~Morinaga, {\em et~al.}, ``Efficient channeling of
  fluorescence photons from single quantum dots into guided modes of optical
  nanofiber,'' {\em Physical review letters} {\bf 109}(6), 063602  (2012).

\bibitem{12}
A.~Goban, K.~Choi, D.~Alton, {\em et~al.}, ``Demonstration of a
  state-insensitive, compensated nanofiber trap,'' {\em Physical review
  letters} {\bf 109}(3), 033603  (2012).

\bibitem{13}
R.~Yalla, M.~Sadgrove, K.~P. Nayak, {\em et~al.}, ``Cavity quantum
  electrodynamics on a nanofiber using a composite photonic crystal cavity,''
  {\em Physical review letters} {\bf 113}(14), 143601  (2014).

\bibitem{14}
S.~Kato and T.~Aoki, ``Strong coupling between a trapped single atom and an
  all-fiber cavity,'' {\em Physical review letters} {\bf 115}(9), 093603
  (2015).

\bibitem{elaganuru2024highly}
B.~Elaganuru, M.~Resmi, and R.~Yalla, ``Highly efficient channeling of single
  photons into guided modes of optical nanocapillary fibers,'' {\em Optical and
  Quantum Electronics} {\bf 56}(5), 893  (2024).

\bibitem{resmi2023efficient}
M.~Resmi, E.~Bashaiah, B.~Das, {\em et~al.}, ``Efficient fiber-coupled single
  photon source using an optical nanofiber tip,'' in {\em Women in Optics and
  Photonics in India 2022},   {\bf 12638}, 107--109, SPIE  (2023).

\bibitem{das2023efficient}
B.~Das, M.~Resmi, E.~Bashaiah, {\em et~al.}, ``Efficient single-mode coupling
  design using a silica/diamond nano-tip with a gold nanoparticle,'' in {\em
  Women in Optics and Photonics in India 2022},   {\bf 12638}, 125--127, SPIE
  (2023).

\bibitem{resmi2024channeling}
M.~Resmi, E.~Bashaiah, and R.~Yalla, ``Channeling of fluorescence photons from
  quantum dots into guided modes of an optical nanofiber tip,'' {\em Journal of
  Optics} {\bf 26}(6), 065401  (2024).

\bibitem{pathak2017fabrication}
A.~Pathak, V.~Bhardwaj, R.~Gangwar, {\em et~al.}, ``Fabrication and
  characterization of tio2 coated cone shaped nano-fiber ph sensor,'' {\em
  Optics Communications} {\bf 386}, 43--48  (2017).

\bibitem{ascorbe2016high}
J.~Ascorbe, J.~Corres, I.~Matias, {\em et~al.}, ``High sensitivity humidity
  sensor based on cladding-etched optical fiber and lossy mode resonances,''
  {\em Sensors and Actuators B: Chemical} {\bf 233}, 7--16  (2016).

\bibitem{resmi2024highly}
M.~Resmi, E.~Bashaiah, S.~Suman, {\em et~al.}, ``Highly efficient coupling of
  single photons using a pair of nanostructures,'' {\em Optical and Quantum
  Electronics} {\bf 56}(8), 1341  (2024).

\bibitem{kaur2021fabrication}
M.~Kaur, G.~Hohert, P.~M. Lane, {\em et~al.}, ``Fabrication of a stepped
  optical fiber tip for miniaturized scanners,'' {\em Optical Fiber Technology}
  {\bf 61}, 102436  (2021).

\bibitem{wu2013optical}
X.~Wu and L.~Tong, ``Optical microfibers and nanofibers,'' {\em Nanophotonics}
  {\bf 2}(5-6), 407--428  (2013).

\bibitem{zhang2024optical}
J.~Zhang, H.~Fang, P.~Wang, {\em et~al.}, ``Optical microfiber or nanofiber: a
  miniature fiber-optic platform for nanophotonics,'' {\em Photonics Insights}
  {\bf 3}(1), R02--R02  (2024).

\bibitem{fang2024parallel}
H.~Fang, Y.~Xie, Z.~Yuan, {\em et~al.}, ``Parallel fabrication of silica
  optical microfibers and nanofibers,'' {\em Light: Advanced Manufacturing}
  {\bf 5}, 1--9  (2024).

\bibitem{praveen2023particle}
P.~Praveen~Kamath, S.~Sil, V.~G. Truong, {\em et~al.}, ``Particle trapping with
  optical nanofibers: a review,'' {\em Biomedical Optics Express} {\bf 14}(12),
  6172--6189  (2023).

\bibitem{wang2024optical}
Z.~Wang, Z.~Chen, L.~Ma, {\em et~al.}, ``Optical microfiber intelligent sensor:
  wearable cardiorespiratory and behavior monitoring with a flexible
  wave-shaped polymer optical microfiber,'' {\em ACS Applied Materials \&
  Interfaces} {\bf 16}(7), 8333--8345  (2024).

\bibitem{15}
N.~Mauranyapin, L.~Madsen, M.~Taylor, {\em et~al.}, ``Evanescent
  single-molecule biosensing with quantum-limited precision,'' {\em Nature
  Photonics} {\bf 11}(8), 477--481  (2017).

\bibitem{suman2024situ}
S.~Suman, E.~Bashaiah, R.~Yalla, {\em et~al.}, ``In situ characterization of
  optical micro/nano fibers using scattering loss analysis,'' {\em Journal of
  Applied Physics} {\bf 135}(12)  (2024).

\bibitem{zaca2018etched}
P.~Zaca-Mor{\'a}n, J.~Padilla-Mart{\'\i}nez, J.~P{\'e}rez-Corte, {\em et~al.},
  ``Etched optical fiber for measuring concentration and refractive index of
  sucrose solutions by evanescent waves,'' {\em Laser Physics} {\bf 28}(11),
  116002  (2018).

\bibitem{tong2003subwavelength}
L.~Tong, R.~R. Gattass, J.~B. Ashcom, {\em et~al.}, ``Subwavelength-diameter
  silica wires for low-loss optical wave guiding,'' {\em Nature} {\bf
  426}(6968), 816--819  (2003).

\bibitem{tong2005self}
L.~Tong, J.~Lou, Z.~Ye, {\em et~al.}, ``Self-modulated taper drawing of silica
  nanowires,'' {\em Nanotechnology} {\bf 16}(9), 1445  (2005).

\bibitem{brambilla2004ultra}
G.~Brambilla, V.~Finazzi, and D.~J. Richardson, ``Ultra-low-loss optical fiber
  nanotapers,'' {\em Optics express} {\bf 12}(10), 2258--2263  (2004).

\bibitem{brambilla2010optical}
G.~Brambilla, ``Optical fibre nanotaper sensors,'' {\em Optical Fiber
  Technology} {\bf 16}(6), 331--342  (2010).

\bibitem{lee2019fabrication}
D.~Lee, K.~J. Lee, J.-H. Kim, {\em et~al.}, ``Fabrication method for ultra-long
  optical micro/nano-fibers,'' {\em Current Applied Physics} {\bf 19}(12),
  1334--1337  (2019).

\bibitem{hoffman2014ultrahigh}
J.~Hoffman, S.~Ravets, J.~Grover, {\em et~al.}, ``Ultrahigh transmission
  optical nanofibers,'' {\em AIP advances} {\bf 4}(6)  (2014).

\bibitem{ward2006heat}
J.~M. Ward, D.~G. O’Shea, B.~J. Shortt, {\em et~al.}, ``Heat-and-pull rig for
  fiber taper fabrication,'' {\em Review of scientific instruments} {\bf 77}(8)
   (2006).

\bibitem{shi2006fabrication}
L.~Shi, X.~Chen, H.~Liu, {\em et~al.}, ``Fabrication of submicron-diameter
  silica fibers using electric strip heater,'' {\em Optics express} {\bf
  14}(12), 5055--5060  (2006).

\bibitem{lazarev2003formation}
A.~Lazarev, N.~Fang, Q.~Luo, {\em et~al.}, ``Formation of fine near-field
  scanning optical microscopy tips. part ii. by laser-heated pulling and
  bending,'' {\em Review of scientific instruments} {\bf 74}(8), 3684--3688
  (2003).

\bibitem{grosjean2007fiber}
T.~Grosjean, S.~S. Saleh, M.~A. Suarez, {\em et~al.}, ``Fiber microaxicons
  fabricated by a polishing technique for the generation of bessel-like
  beams,'' {\em Applied optics} {\bf 46}(33), 8061--8067  (2007).

\bibitem{kbashi2012fabrication}
H.~J. Kbashi, ``Fabrication of submicron-diameter and taper fibers using
  chemical etching,'' {\em Journal of Materials Science \& Technology} {\bf
  28}(4), 308--312  (2012).

\bibitem{liao2010suspended}
M.~Liao, C.~Chaudhari, X.~Yan, {\em et~al.}, ``A suspended core nanofiber with
  unprecedented large diameter ratio of holey region to core,'' {\em Optics
  Express} {\bf 18}(9), 9088--9097  (2010).

\bibitem{bashaiah2024fabrication}
E.~Resmi~M, Bashaiah and R.~Yalla, ``Fabrication and characterization of
  optical nanofiber tips,'' {\em Journal of Nanophotonics} {\bf 18}(2),
  026007--026007  (2024).

\bibitem{huo2006fabrication}
X.~Huo, S.~Pan, and S.~Wu, ``Fabrication of optical fiber probe nano-tips by
  heated micro-pulling combined with static chemical etching,'' in {\em 2006
  1st IEEE International Conference on Nano/Micro Engineered and Molecular
  Systems},  254--257, IEEE  (2006).

\bibitem{ren2007preparation}
H.~Ren, C.~Jiang, W.~Hu, {\em et~al.}, ``The preparation of optical fibre
  nanoprobe and its application in spectral detection,'' {\em Optics \& Laser
  Technology} {\bf 39}(5), 1025--1029  (2007).

\bibitem{birks1992shape}
T.~A. Birks and Y.~W. Li, ``The shape of fiber tapers,'' {\em Journal of
  lightwave technology} {\bf 10}(4), 432--438  (1992).

\bibitem{nayak2018nanofiber}
K.~P. Nayak, M.~Sadgrove, R.~Yalla, {\em et~al.}, ``Nanofiber quantum
  photonics,'' {\em Journal of Optics} {\bf 20}(7), 073001  (2018).

\bibitem{politi2009integrated}
A.~Politi, J.~C. Matthews, M.~G. Thompson, {\em et~al.}, ``Integrated quantum
  photonics,'' {\em IEEE Journal of Selected Topics in Quantum Electronics}
  {\bf 15}(6), 1673--1684  (2009).

\bibitem{hale1994demonstration}
Z.~Hale and F.~Payne, ``Demonstration of an optimised evanescent field optical
  fibre sensor,'' {\em Analytica chimica acta} {\bf 293}(1-2), 49--54  (1994).

\bibitem{pace2004refractive}
P.~Pace, S.~T. Huntington, K.~Lyytik{\"a}inen, {\em et~al.}, ``Refractive index
  profiles of ge-doped optical fibers with nanometer spatial resolution using
  atomic force microscopy,'' {\em Optics Express} {\bf 12}(7), 1452--1457
  (2004).

\end{thebibliography}
\bibliographystyle{spiejour}   % makes bibtex use spiejour.bst

{}
\author[a]{Shashank Suman}
\author[a]{Resmi M}
\author[a]{Bratati Das}

\vspace{2ex}\noindent\textbf{Elaganuru Bashaiah} is a PhD scholar at the University of Hyderabad, India. He received his MSc degree in Physics from Vikrama Simhapuri University, India in 2016. He is the author of 5 articles including journal and conference papers. 

\vspace{2ex}\noindent\textbf{Shashank Suman} is a Research Fellow at the University of Hyderabad, India. He received his 5-year Integrated M.Sc. in Physics UM-DAE Center for Excellence in Basic Sciences, India in 2023. 

\vspace{2ex}\noindent\textbf{Resmi M} is a PhD scholar at the University of Hyderabad, India. She received her MSc degree in Physics from University of Madras, India in 2016. She is the author of 5 articles including journal and conference papers.

\vspace{2ex}\noindent\textbf{Bratati Das} is a PhD scholar at the University of Hyderabad, India. She received her MSc degree in Physics from Jawaharlal Nehru University, India in 2018.

\vspace{2ex}\noindent\textbf{Ramachandrarao Yalla} is an assistant professor at the University of Hyderabad, India. He received his MSc degree in Physics from the University of Hyderabad, India in 2008, and his PhD degree in Physics from the University of Electro Communications, Japan in 2012. He is the author of more than 18 journal papers including PRL papers. His current research interests include quantum optics, nanophotonics and optoelectronic systems. He is a member of SPIE.

\end{spacing}
\end{document}